\documentclass{jfm}
\usepackage{graphicx}
\usepackage{gensymb,amsmath,amssymb,hyperref}

\usepackage{epstopdf, epsfig}
\usepackage{xcolor}
\definecolor{darkblue}{rgb}{0.2,0.2,0.6}
\hypersetup{colorlinks,breaklinks,
            linkcolor=darkblue,urlcolor=darkblue,
            anchorcolor=darkblue,citecolor=darkblue}
\usepackage{comment}

\newcommand{\dev}{}

\shorttitle{Flat disc slams on water: Air Entrapment and Impulse transfer}
\shortauthor{U. Jain, P. Vega-Mart\'{i}nez and D. van der Meer}

\title{Air Entrapment and its effect on Pressure Impulses in the slamming of a Flat Disc on Water}

\author{Utkarsh Jain\aff{1}
  \corresp{\email{u.jain@utwente.nl}},
  Patricia Vega-Mart\'{i}nez\aff{2} \and Devaraj van der Meer\aff{1}}

\affiliation{\aff{1}Physics of Fluids Group and Max Planck Center Twente for Complex Fluid Dynamics, MESA+ Institute and J. M. Burgers Centre for Fluid Dynamics, University of Twente, P.O. Box 217, 7500AE Enschede, The Netherlands
\aff{2}Fluid Mechanics Group, Universidad Carlos III de Madrid, Legan\'{e}s, Spain}

\begin{document}

\maketitle

\begin{abstract}
The presence of ambient air in liquid-slamming events plays a crucial role in influencing the shape of the liquid surface prior to the impact, and the distribution of loads created upon impact. We study the effect of trapped air {on impact loads} in a simplified geometry, by slamming a \dev{horizontal} flat disc \dev{onto} a stationary water bath \dev{at a well-controlled velocity}. We show how air trapping influences pressure peaks at different radial locations on the disc, how the pressure impulses are affected, and {how} local pressure impulses differ from \dev{those obtained from area-integrated (force)} impulses at impact. \dev{More specifically, we find that the air layer causes a gradual buildup of the load before the peak value is reached, and show that this buildup follows inertial scaling.}
{Further, the same localised pressure impulse at the disc centre are found to be lower than the corresponding (area-integrated) force impulse on the entire disc. While the (area-integrated) force impulses \dev{are close to} the classical result of \citet[section 6.10]{batchelor1967introduction} and \citet{glasheen1996vertical}, the localised pressure impulses at disc center, where the trapped air layer is at its thickest, are found to lie closer to the theoretical estimation by \citet{peters2013splash} for an air-cushioned \dev{impact}.}
\end{abstract}


\section{Introduction}

Situations involving the impact of a liquid on solid, or vice-versa are commonly found in several industrial applications. Some notable examples that occur on a large scale are hull slamming during the (re-)entry of a ship into water \citep{kapsenberg2011slamming}, the landing of sea-planes or spacecraft \citep{abrate2011hull}, wet-deck slamming in ships or off-shore structures \citep{smith1998slam,faltinsen2000hydroelastic,faltinsen2004slamming}, and wave impacts on harbours and sea-walls  \citep{peregrinereview}, and during sloshing \citep{dias2018slamming}. Liquid-solid impact events are associated with large loads that are created primarily due to the added mass effect: upon impact, a certain mass of liquid is rapidly accelerated towards the velocity of the impacting object. However in all situations, the surrounding air can have a very non-trivial influence on the generated load, and its distribution on the solid structure \citep{dias2018slamming, bogaert2010sloshing}. One way this occurs is via aeration of the liquid phase \citep{bredmose2015violent, ma2016pure}. More often, the air that is trapped in between the liquid and the solid {affects the shape of the liquid surface \citep{hicks2012air}, and} also plays a crucial role in affecting the loading on the solid phase \citep{bredmose2009violent, peregrine1996effect, wood2000wave, hattori1994wave, ERMANYUK2005345}. {In this work, we focus on the role of trapped air}. Its role in influencing the pressures can be benign, such as by mitigating pressure singularities and distributing the loads over a larger surface. On the contrary, the trapped air can also increase the {duration of the} load{ing, and cause additional damage to a solid structure} by {an oscillating air pocket} or a cavitation implosion {of the trapped air}.

Here we experimentally study the local pressures and forces produced on a disc that is impacted onto a water bath. Prior to the impact, the disc creates an air flow that results in it entrapping air on the impacting side of its own accord \citep{verhagen1967impact, ujthesis}. We describe the setup and the air trapping on the disc in section \ref{sec:setup}. The trapped air layer has been shown to cause a small reduction in the peak pressure \citep{chuang1966experiments} and also affect its distribution \citep{okada2000water, todter2020experimentally} over the impacting face. In section \ref{sec:pressuressection}, we report the measurements of local pressures at the centre of the disc and near its edge (see figure \ref{fig:forceimpulsesetup}), and how the trapped air influences peak pressures. The discussion in this section is further complemented in section \ref{sec:loadcellsection} by measuring the total impact force on the assembly with a load cell. The peak values of both are found to be very sensitive to the temporal resolution of the equipment{, something that is known from liquid-slamming literature over decades.} \dev{This is connected to the fact that theoretically, in the absence of air cushioning, the moment of impact creates a temporal pressure singularity at the point of contact on the solid.}

{\dev{It is known} that \dev{therefore time-integrated pressure signals, also known as} pressure impulses\dev{,} are a much more faithful indicator of the loading intensity \citep{bagnold1939interim, hattori1994wave, partenscky1989dynamic, bullock2007violent}.}
\dev{Consequently, we} 
compute impulses from our measurements, and find \dev{them to be significantly more well-behaved in the region of the} peak pressures and forces.  \dev{We show that the buildup of the load before the peak value is reached is mediated by the air layer, and follows inertial scaling for both pressure and force impulses. Turning to the impulse accumulated during the impact peak, the (area-integrated) force impulses are found to lie very close to the classical result of \citet[section 6.10]{batchelor1967introduction} and \citet{glasheen1996vertical}, while the local pressure impulses at the disc centre lie along the air-cushioned impact impulse calculation of \citet{peters2013splash}.}

\section{Setup}\label{sec:setup}

\begin{figure}
    \centering
    \includegraphics[width=.99\linewidth]{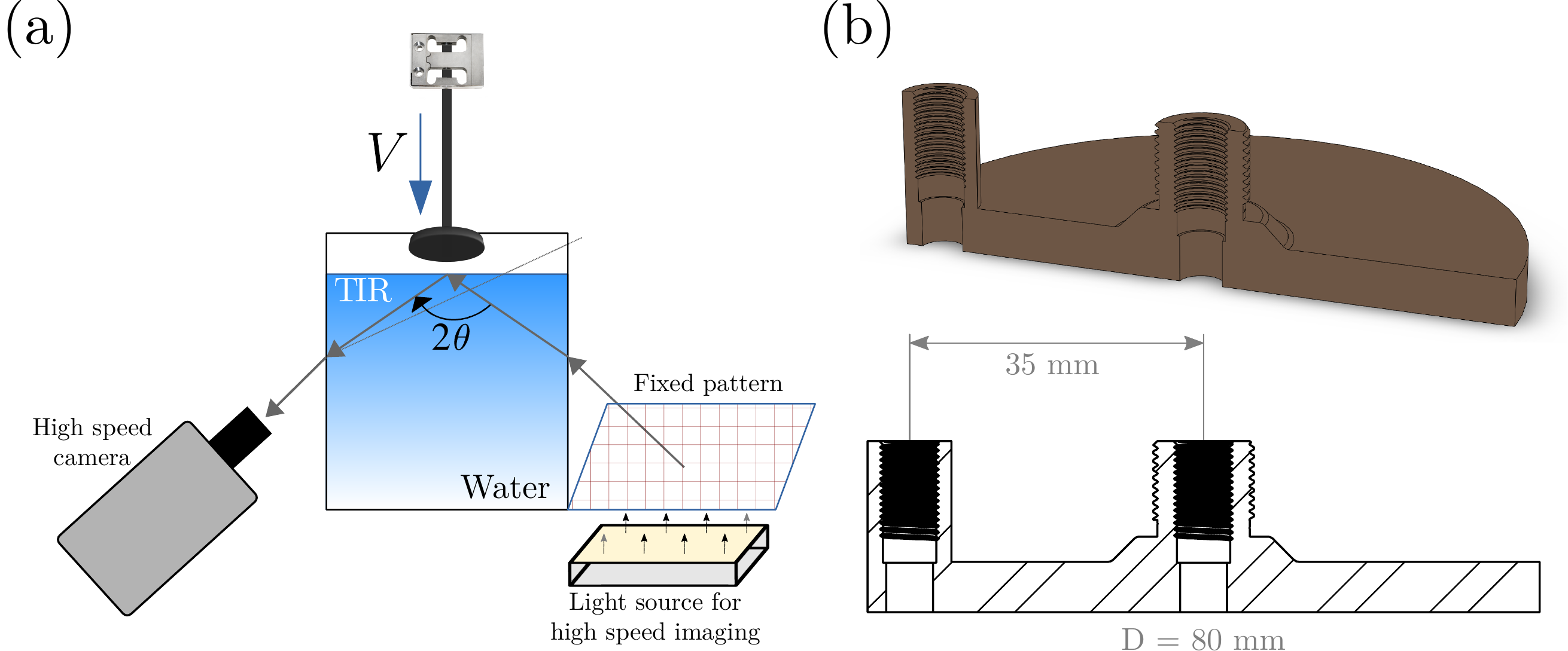}
    \caption{(a) Schematic of the setup. A linear motor is used to impact a disc on water surface with constant velocity. Using total internal reflection (TIR), the water surface can be used as a mirror, and its behaviour observed when subjected to deflections. A load cell is installed along the rod to measure the overall force experienced in the vertical direction. (b) Design (upper figure) and dimensions (lower figure) of an 80 mm wide disc on whose surface local impact pressures are measured. {The two identical sensors have a circular sensing area with a diameter of 5.5 mm. Their mounting locations - at the disc's center, and close to its edge, are shown in panel (b).}}
    \label{fig:forceimpulsesetup}
\end{figure}

The {experimental installation} consists of a reservoir filled with water, which is contained in a tank of area 50 cm $\times$ 50 cm. The water depth is kept fixed at 30 cm. We impact flat discs on the water bath such that their impacting surface is parallel to the stationary water surface. The disc's motion across the free surface is controlled using a linear motor, with a position accuracy of approximately 0.6 $\mu$m. The assembly's motion is programmed such that it achieves a specified velocity when in the middle of its available stroke. Concerning the disc, the stroke is programmed such that it attains the constant, programmed-for velocity $V$, several diameters before it reaches the free surface. This constant velocity is maintained as the disc plunges roughly the same distance into the target pool. $V$ is varied from 0.05 -- 3 m/s.

The free surface of water is \dev{both} visualised \dev{directly, and} using the total internal reflection (TIR) setup as shown by the schematic in figure \ref{fig:forceimpulsesetup}a \citep{tirdmanuscript,jainkhPRF}. When the disc and water surface are in touch, those sections of the free surface that wet the disc cease to reflect the reference pattern and effectively disappear.

We install pressure sensors on an 80 mm wide disc made of steel to measure pressures on its impacting surface. Vaseline is used to seal any gaps between the sensors and the mounting sites on the disc. The sensors used were Kistler 601CAA dynamic piezoelectric transducers, with a diaphragm width of 5.5 mm. {All the pressure measurements reported throughout the paper were done with a single disc of diameter 80 mm. The sensors together cover an area of approximately 0.95\% of the disc's impacting surface.} They were flush mounted on the disc as per the manufacturer's instructions. A sketch showing the dimensions of the disc and positions of the sensors is shown in figure \ref{fig:forceimpulsesetup}(b). Pressure measurements were made at an acquisition frequency of 200 kHz, {where the sensors have a natural frequency larger than 215 kHz.} Additionally we also measure the overall force on the disc that is generated at impact using {analog} strain gauge load cell{s (FUTEK LSM300). Over the range of force measurements performed, the load cell with a suitable maximum loading capacity was used - with 200 and 500 lbs, with natural frequencies of 1773 and 2499 Hz respectively}. {The load cells were} calibrated using known weights, and measurements done {using an analog amplifier (FUTEK IAA200) with current output} at an acquisition rate of 25 kHz. In this case, disc radii $R$ are varied from 1--8 cm. {The pressure measurements were done with a single disc of stainless steel (RVS 316), with density 8.03 g/cm$^3$. Force measurements were done with a number of steel discs (RVS 316), aluminium (6082 aluminium alloy, with density 2.71 g/cm$^3$) and 3D printed plastic (with density 1.13 g/cm$^3$). Their tensile properties appear to be irrelevant even for the short timescales over which peak forces are resolved (see figure \ref{fig:force1}).} {Natural frequencies of the impacting assemblies attached to the respective load cells were measured, and are mentioned alongside time series where shown in the paper.}

\section{\dev{Air film entrapment and spatially localised loading}}\label{sec:pressuressection}

\subsection{\dev{Air film entrapment}}

\begin{figure}
    \centering
    \includegraphics[width=.99\linewidth]{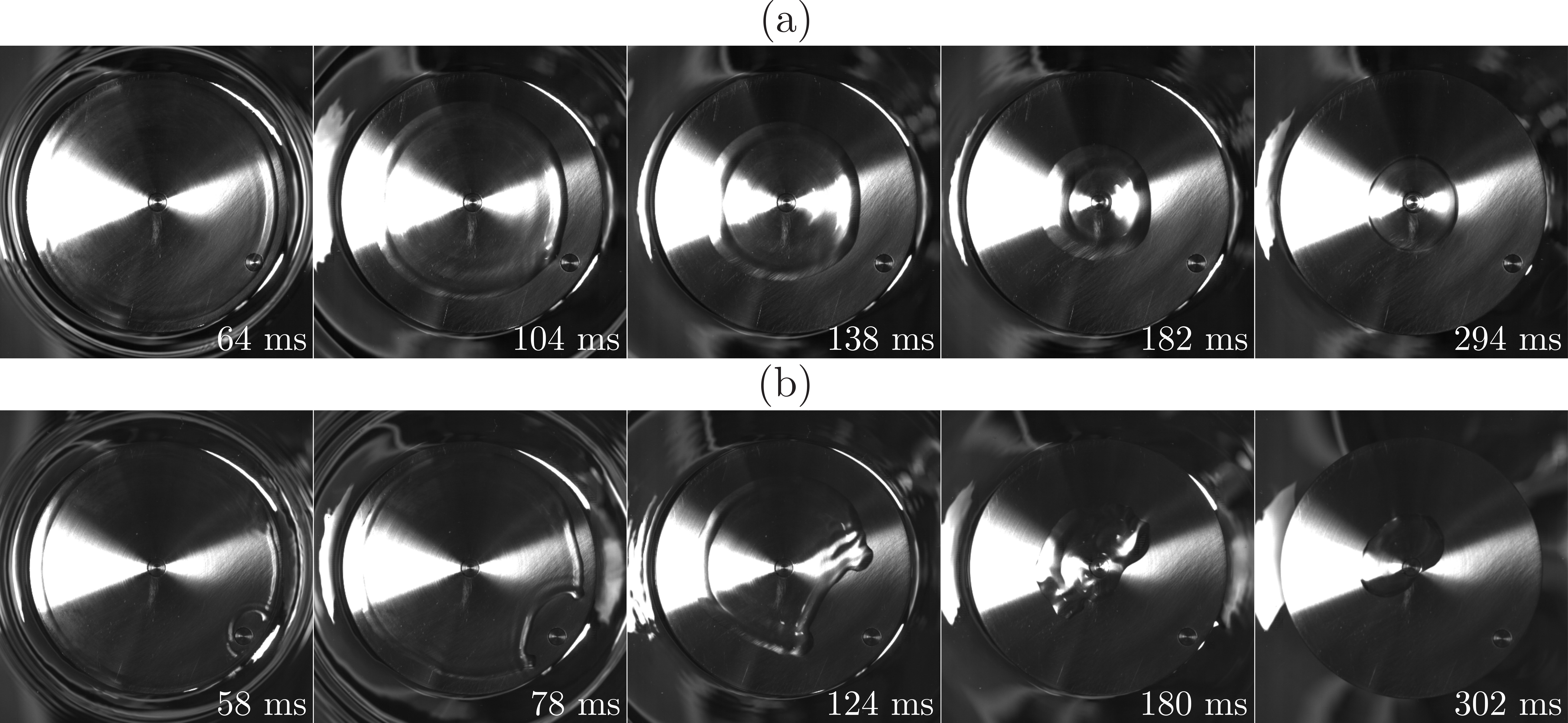}
    \caption{The trapping of an air film on a water-entering disc is directly seen from below the water bath at $V = $ 0.06 m/s in panels (a), and at 0.1 m/s in panels (b). The disc shown is 8 cm wide (as shown in figure \ref{fig:forceimpulsesetup}(b)).The time labels are centred at 0 ms when the disc makes it first contact with water. The trapped air film contracts inwards due to surface tension of water air interface. The process is also shown in Movies 1 and 2. {The film's retraction dynamics are discussed in appendix \ref{sec:appairfilmretraction}.}}
    \label{fig:airfilmcollapse}
\end{figure}

\begin{figure}
    \centering
    \includegraphics[width=.99\linewidth]{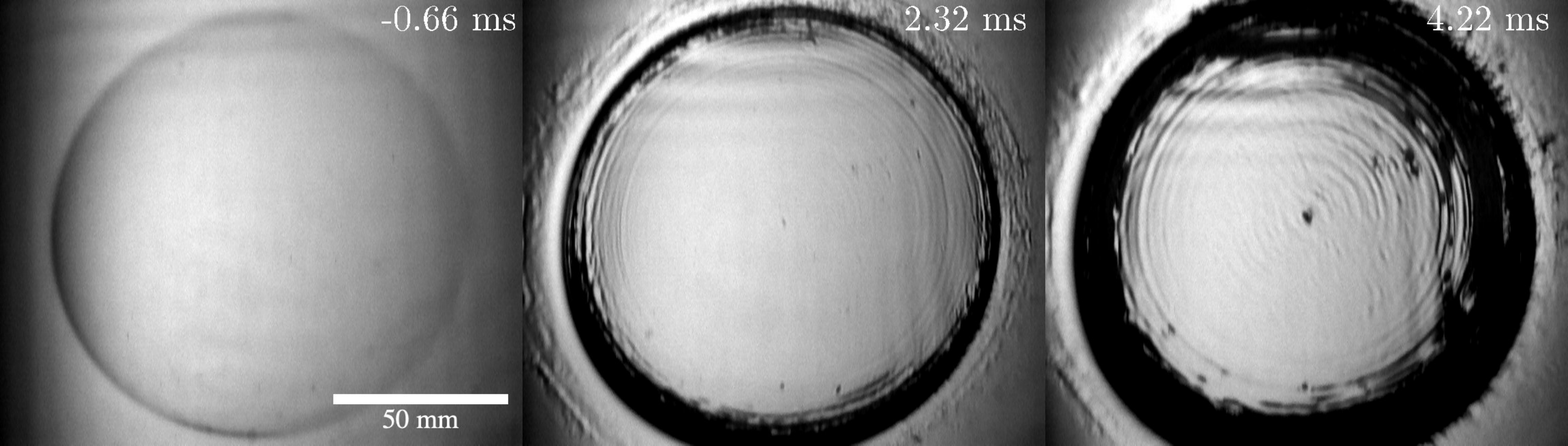}
    \caption{Snapshots of the free surface during the impact of a {12 cm wide disc at 0.5 m/s}, {recorded at 50k fps} using the TIR setup, as shown in figure \ref{fig:forceimpulsesetup}(a), imaging the free surface from below. {The free surface is pushed down before the disc makes contact with it, which is evident from the first panel on the left.} The time labels shown at top left corner are centred about $t=0$ at the instant where the disc makes first contact with the deformed water surface. An air layer is trapped on the disc at impact. It collapses inwards as the disc plunges further into the liquid bath \dev{(centre panel) and is finally punctured at the center by the liguid (right panel)}. \dev{The puncturing process is shown in Movie 4 and discussed in further detail in appendix \ref{sec:appairfilmpuncture}.} More examples using different parameters are shown in movies 5 and 6.}
    \label{fig:12cmpostimpact}
\end{figure}

As the disc approaches the water surface with a finite speed, there is an intervening air layer trapped between the two. The air in this layer is squeezed out such that its flow results in a radial pressure gradient. These pressure changes are sufficiently large to deform the water surface before the disc makes any contact with it. Thus the water surface is made aware of the approaching disc in advance; at the point that lies directly under the disc centre, the liquid surface is pushed down by \dev{the} stagnation pressure \dev{in the air phase}. Conversely, under the disc edge, a high hydrodynamic pressure \dev{in the air} creates a suction which destabilises the water surface to be sucked upwards towards the disc \citep{ujthesis, jainkhPRF}. As a result, at the first moment \dev{that} the disc \dev{makes} contact with the water surface, the water-air \dev{interface} under the disc edge is vertically separated from that under the disc centre by approximately 200 $\mu$m. This `air-cushioning' effect on the free surface invariably creates conditions to trap an air layer under the disc at its first contact with the liquid surface, as is shown in figures \ref{fig:airfilmcollapse} and \ref{fig:12cmpostimpact}.

At low \dev{impact velocity} $V$, such as shown in figure \ref{fig:airfilmcollapse} at 0.06 m/s and 0.1 m/s, \dev{after entrapment} the air layer merely contracts inwards on account of high surface tension of the air-water interface. \dev{It is observed} that, as the surrounding water displaces air, its perimeter stays remarkably stable and consistent in shape. This is due to the interface being stabilised by Saffman-Taylor mechanism, which is only unstable to perturbations when a less viscous liquid displaces a more viscous one. Note how at a slightly higher $V$ of 0.1 m/s in figure \ref{fig:airfilmcollapse}(b) the air film is briefly ruptured as it passes over the sensor at the disc edge, but then continues to contract inwards preserving the interface shape to remain as a bubble at the disc centre. {The inward retraction of the film is due to the high surface tension of air-water interface, and occurs with a constant velocity \dev{as is discussed in more detail in appendix \ref{sec:appairfilmretraction})}.} Our observations of how such a trapped air bubble behaves are consistent with the observations of \citet{mayer2018flat}, who used a flat rectangular plate as the impactor.

At \dev{a somewhat} higher impact velocity \dev{$V = 0.5$ m/s} (figure \ref{fig:12cmpostimpact}) \dev{the trapped air film is shown at the first moment of impact in the left panel, using the TIR technique shown in figure \ref{fig:forceimpulsesetup}(a). With this technique, all parts of the water surface that turn black indicate where the free surface has disappeared either due to wetting the disc, or due to being ejected upward out of the initially quiescent liquid surface's plane. The moment when the disc makes first contact with the water surface can be identified when a section of the reflecting water surface disappears (i.e., turns black) by wetting the disc. In the central panel, it is observed how the air layer initially starts to collapse towards the disc centre while the disc plunges further into the liquid bath, until, in the right panel, the air bubble is punctured in the center due to the high pressure that occurs in the stagnation point that has formed in the liquid. This particular observation is further addressed in appendix \ref{sec:appairfilmpuncture} where we present the measured puncture times. The current and additional examples of this process can be seen in Movies {4--6}.}

\dev{For sufficiently high impact velocities, after the air layer is punctured it is observed to be expelled from below the disc. This happens when the hydrodynamic forces that push the air outwards are stronger than the capillary forces that try to keep the air layer together. }

{\dev{Taking a look at the dimensionless numbers that may play a role during the impact, we note that the} growth of the small cavity of thickness of order $100$ microns, that is formed prior to impact on the free surface, is mainly driven by the stagnation pressure \dev{in the air phase at} the disc's symmetry axis. The formation of such a cavity is resisted by both surface tension and inertia of the liquid. The Weber numbers $\mathit{We} =  \rho_{w} V^2 R/\sigma$ that concern both the entrapment of the air film, and \dev{the subsequent impact}, 
range between approximately \dev{$1$} and $10^4$ (using $R=1-8 $ cm and $V = 0.05 - 3$ m/s).}
{The Froude number $\mathit{Fr} = V^2/gR$ lies in the range between $0.006$ and $92$. Note that in the pre-impact stage, the liquid inertia limits free surface deformation. In this particular stage prior to impact, the effective Froude number is $\mathit{Fr}\rho_{\text{gas}}/\rho_{\text{water}}$, which is much smaller than the \dev{uncorrected Froude and} Webers number at impact. Shortly after impact when a large slamming force is created on the impactor, the liquid experiences much larger acceleration than gravity, making the Froude number irrelevant \citep{peters2013splash,mayer2018flat,iafratikorobkin2004}.}

\dev{In the next subsection, we will discuss the pressures that are measured during impact by the two pressure sensors in the $R=4.0$ cm steel disk of figure~\ref{fig:forceimpulsesetup}. Before starting this discussion, we want to however stress the difference in time scales in the buildup of the pressure on the one hand ($\ll 1$ ms) and that of the contraction ($\sim 100$ ms) and the puncturing ($> 1$ ms) and sometimes subsequent expulsion ($\sim 10$ ms) of the entrapped air layer on the other. This implies that, during pressure buildup, the air film can be viewed as largely stationary and covering most of the disc.}

\subsection{Pressure measurements}
\begin{figure}
    \centering
    \includegraphics[width=.999\linewidth]{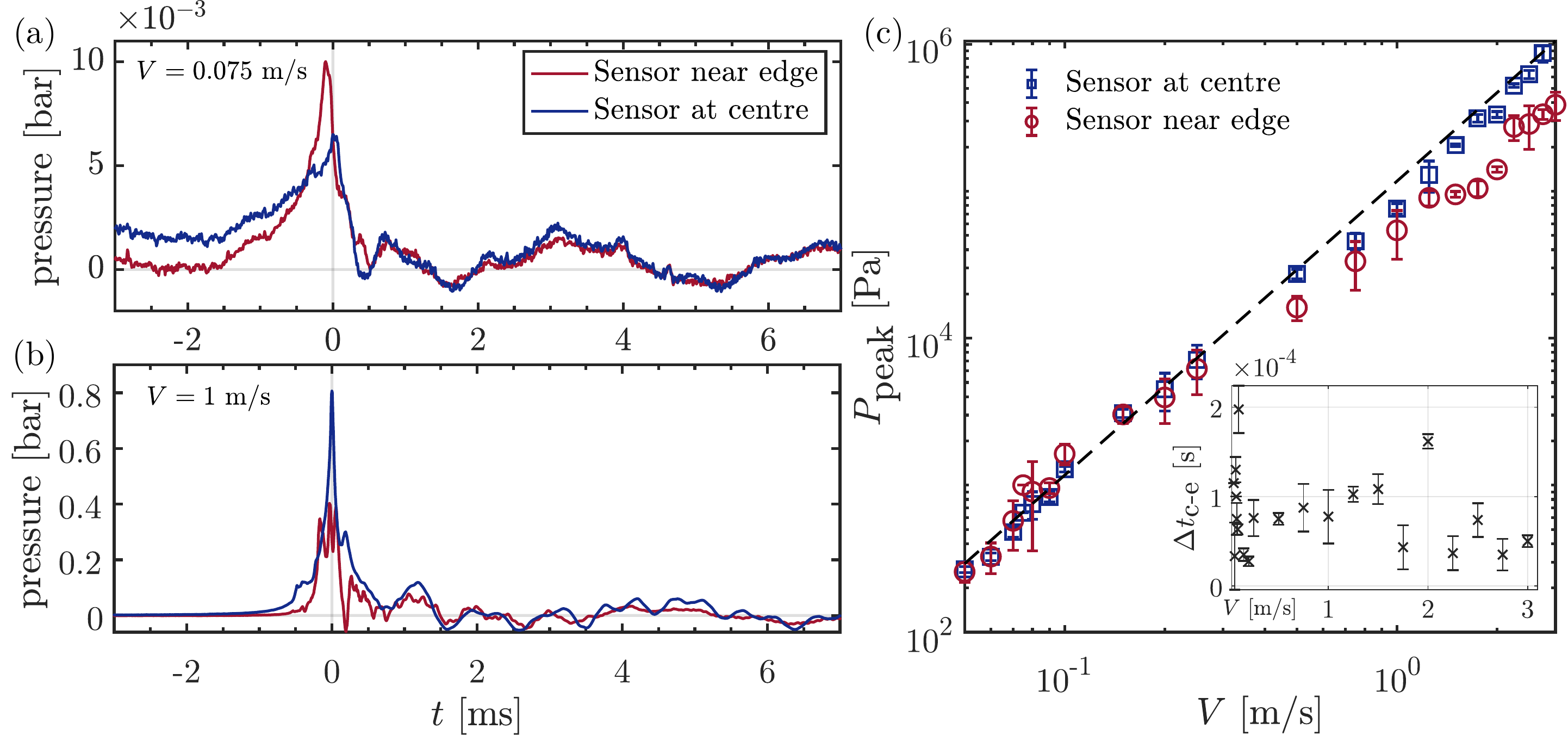}
    \caption{Time series of pressures measured at the two locations on an 80 mm wide disc (see figure \ref{fig:forceimpulsesetup}b) when it impacts on a deep liquid bath at a velocity of (a) 0.075 m/s and (b) 1.0 m/s. The signals are centred about $t=0$ at the instant when the central pressure reaches its peak. (c) Peak pressures measured at two locations on the same disc are plotted over a range of impact velocities. Error bars show the standard deviation of peak pressures over 3--6 repetitions of the experiment at each impact velocity. The peak pressures at the centre are found to scale quadratically with $V$, whereas those near the edge of the disc are found to be slightly below those at the centre. {The impact peak is always attained first at the edge sensor, irrespective of where a higher peak pressure is attained. The inset of panel (b) highlights this by showing the time difference $\Delta t_{\text{c-e}} = t_\text{center} - t_\text{edge}$ between the times $t_\text{center}$ and $t_\text{edge}$ at which the peak pressure is attained at the center and the edge, respectively.}}
    \label{fig:peakpressuresquadscaling}
\end{figure}

The effect of impact on pressure characteristics {is} found to be typical of solid-liquid impact events in general \citep{peregrinereview, kapsenberg2011slamming}. The first contact between a transducer on the disc {edge} and water surface creates a large pressure spike (figure \ref{fig:peakpressuresquadscaling}(a,b)), lasting for a short duration $\sim \mathcal{O}(10^{-4} \text{ s})$. Note from figure \ref{fig:peakpressuresquadscaling}(a,b) how in examples shown for both a low velocity impact at $V = $ 0.075 m/s, and a larger velocity of $1$ m/s, the pressure at disc centre starts to grow prior to that at the edge. This is due to a 
stagnation pressure building up in the liquid at the disc centre, whose magnitude is of the order $\rho_w V^2$. {This can be expected, since the free surface under the disc responds to the pressure buildup before the disc comes into contact with the water. The magnitude of such a pressure buildup will exist in an approximate balance with the high pressures needed to accelerate the fluid via the cushioning layer. Another important observation from panel (a) is that while the time taken for the air film to retract from the edge sensor to the one at centre is $\approx 185$ ms (see also figure \ref{fig:airfilmcollapse}, {appendix \ref{sec:appairfilmretraction}}, and supplementary movies 1 and 2), the time gap between the pressure peaks at the two sensors is of the order $0.1$ ms {(also shown in inset figure \ref{fig:peakpressuresquadscaling}(b))}. We can conclude that the both the pressure buildup, and the attainment of the peak pressure at centre, are certainly mediated by the air cushion.}

{While it is clear that the first contact at the disc's edge creates the first peak pressure, it is not immediately clear where the bubble's outer perimeter lies, and \dev{what shape it has} when the peak pressure at the center (a mere $\sim \mathcal{O}(10^{-4} \text{ s})$ later, inset figure \ref{fig:peakpressuresquadscaling}(b)) is registered. The time delay \dev{$\Delta t_{c-e}$ between the peaks at the edge and center} is much shorter than \dev{the time span in} which the trapped air retracts ($\sim \mathcal{O}(10^{-1} \text{ s})$, also see figure \ref{fig:airfilmretraction}), or over which it ruptures and fragments at higher velocities ($\sim \mathcal{O}(10^{-3} \text{ s})$, also see figure \ref{fig:onsetofpuncture}).} \dev{In addition, the impact velocity does not clearly affect $\Delta t_{c-e}$.}

A{n oscillating} component in the pressure signals with a period of $\sim 4$ ms is visible at all times in panel (a), and after the impact in panel (b) in figure \ref{fig:peakpressuresquadscaling}. This arises from structural oscillations in the disc-rod assembly, and is also present in the data from the load cell (figure \ref{fig:force1}). {The natural frequency of this impactor assembly (with masses of rod and disc combining to approximately 502 grams) was measured with the load cell to be 236.8 $\pm$ 30.5 Hz, corresponding to a period of approximately 4.2 ms}.

Peak pressures measured from impacts done at a range of velocities are plotted in figure \ref{fig:peakpressuresquadscaling}(c). Multiple repetitions of the experiment at each velocity were performed to quantify reproducibility. At low velocities ($V \lesssim 0.5$ m/s), pressures at the edge of the disc were found to be higher than those at the disc's centre. The opposite however was observed in impacts at higher velocities ($V \gtrsim 0.5$ m/s). Thus, while at \emph{low} velocities the impact pressures at the disc's centre are actively `cushioned' by the trapped air film, for $V \gtrsim 0.5$ m/s, the air film appears to play a diminished role 
in reducing the peak pressure at the disc centre. Note that peak impact pressures at the disc centre are approximately 100 times larger than the stagnation pressure $\rho_w V^2$ in water \dev{and that they are found to scale quadratically with $V$, as expected for an inertial impact process} (see figure \ref{fig:peakpressuresquadscaling}(c)). \citet{okada2000water} performed impact tests using a flat plate, and also found higher peak pressures closer to its centre. We thus designate impact velocities $\gtrsim 0.5$ m/s as `slamming' velocities throughout the present work. At slamming velocities, {the initially trapped air film does not persist - either getting punctured by the stagnation pressure ($\sim \mathcal{O} (10^{-3} \text{ s})$ after impact, see appendix \ref{sec:appairfilmpuncture}), or fragmenting prior to being} violently expelled out from under the disc ($\sim \mathcal{O} ( 10^{-2} \text{ s})$ after impact, Movie 3). {It is to be emphasised here that \dev{the threshold $V \approx  0.5$ m/s (corresponding to $\mathit{We} \approx 278$) is only intended to differentiate between low-speed impacts (where the entrapped air bubble contracts) and slamming impacts (where the entrapped air bubble is punctured and expelled) in the case of the $R=4.0$ cm disk.}}

\subsection{Localised pressure-impulses during impact}

The most elementary measure of the intensity of an impact event is the peak pressure that is measured on the solid structure. However, in most circumstances, this peak pressure can have a large variance between tests performed at the same parameters. Since rise times of the initial impact peak can be very short $\sim \mathcal{O}(10^{-5} \text{ s})$, one reason for the variance in peak pressures is often due to limited time resolution of the sensors. Another reason very often encountered is the inconsistent role played by the ambient air that is trapped at impact. Since the first such work by \citet{bagnold1939interim} on studying the effect of trapped air on peak pressures, several later works \citep{denny1951further, chanmelville88, peregrinereview} have also identified that the trapped air can play an important role during the production of the initial large pressure peak. \citet{richert1969experimental}, \citet{partenscky1989dynamic} and \citet{hattori1994wave} found from wave impacts on walls, that the most severe pressures were produced when there were several small air bubbles, or a very thin `lens shaped' air pocket trapped along the wall. It is a general consensus that pressure impulse, or the pressure integrated over duration of the peak, is a better predictor of the amount of local damage that is caused by an impact event on the solid structure \citep{bullock2007violent,bredmose2009violent,hull2002investigation,kirkgoz1990experimental,dias2018slamming,peregrinereview}. Indeed, pressure impulse is also found to be a much more reproducible measure of the intensity of impact than the peak value in itself \citep{bagnold1939interim,dias2018slamming,denny1951further,chanmelville88,peregrinereview,bullock2007violent}.

\begin{figure}
    \centering
    \includegraphics[width=.99\linewidth]{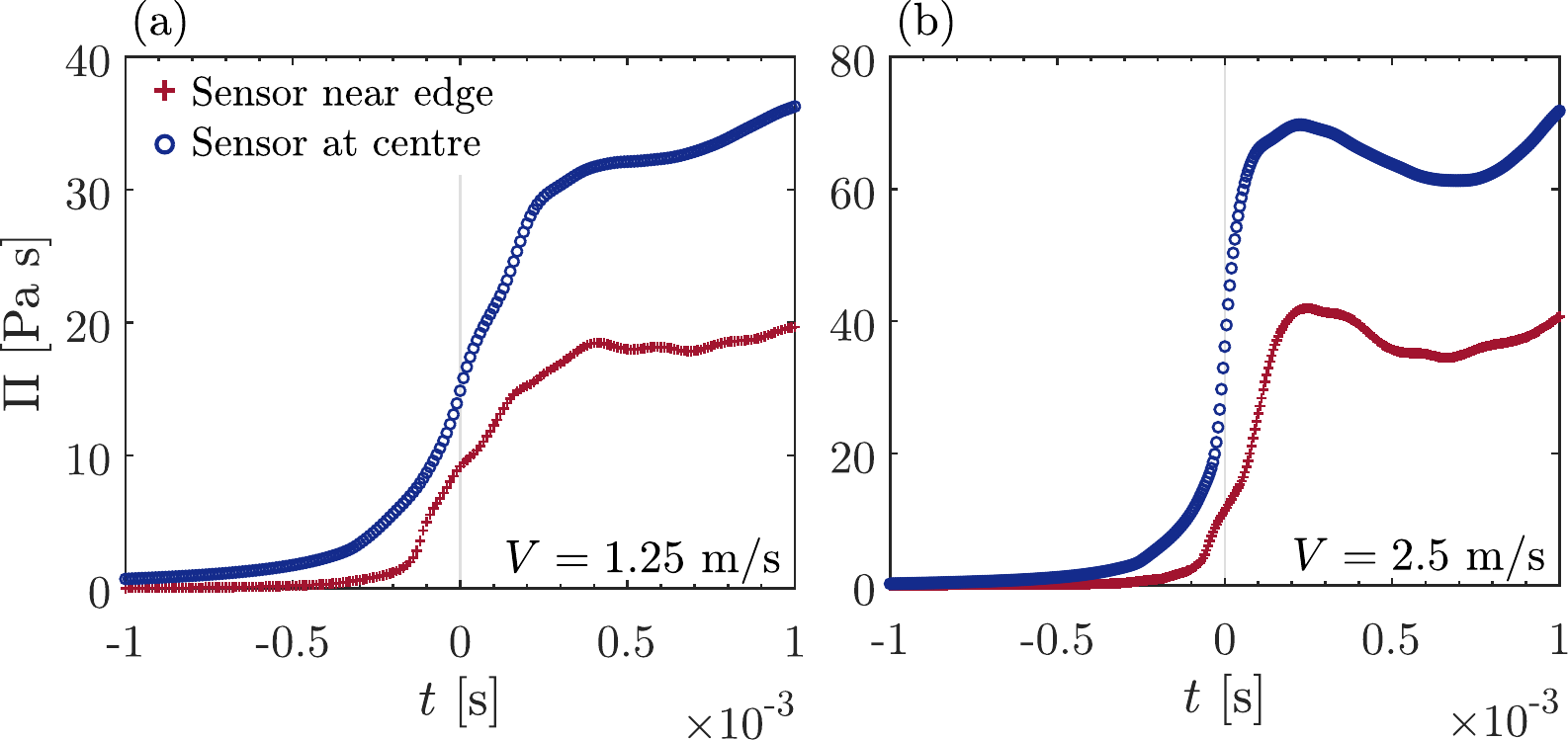}
    \caption{Pressure impulses $\Pi$ measured from experiments at (a) $V = $ 1.25 m/s and (b) $V = $ 2.5 m/s are plotted as time series. Experiments were done with the sensor arrangement as shown in figure \ref{fig:forceimpulsesetup}b. As before, all time series are centred about $t=0$ at moment when the pressure at disc centre reaches its maximum.}
    \label{fig:pressuresensorimpulsesdim}
\end{figure}

Thus we calculate the pressure impulses $\Pi(t)$ from the measurements done at slamming velocities. They are computed by numerically integrating the pressure signal from a time before it starts to rise $t_0$, to arbitrarily later times as \begin{equation}
    \Pi(t) = \displaystyle\int_{t_0}^{t} p(t') dt'.
\end{equation}
Consistent with all the pressure measurements, the time-coordinate of all pressure-impulse results are again centred about $t=0$, at the instant when the central pressure reaches its maximum. This way of presenting the results allows one to focus on the early-time rise in impulse until the trapped air film collapses at the disc's centre.

Pressure impulses measured at $V =$ 1.25 and 2.5 m/s are plotted in figure \ref{fig:pressuresensorimpulsesdim}. At the disc centre, $\Pi(t)$ starts to rise earlier than at the disc's edge. This can be ascribed to an observation made earlier in figure \ref{fig:peakpressuresquadscaling}(a), where it was seen that the pressure at the disc centre starts to rise before than that at the edge. The difference in how impulse at the edge grows compared to that at the centre occurs due to different behaviour of air flow in the two regions at impact. At the moment of impact, there is a trapped air film whose thickness is maximum under the disc centre. The opposite holds true at the disc edge, where the contact first happens. {At slamming velocities, after the peak impact pressure is attained, the air film collapses.} Immediately after this, the air that was initially trapped is fragmented and expelled from under the disc. As its escape happens from under the disc edge, fast air flow, followed by fast liquid flow, interfere with the pressure reading. That the impulse shown in figure \ref{fig:pressuresensorimpulsesdim} initially accumulates more smoothly at the centre than at the disc's edge suggests that impact at the disc centre is cushioned by the air layer. Indeed, the presence of air also prolongs the duration of the initial impact peak, which contributes to the impulse at the centre growing to larger values than that at disc edge \citep{iafrati2008hydrodynamic}.

\begin{figure}
    \centering
    \includegraphics[width=.65\linewidth]{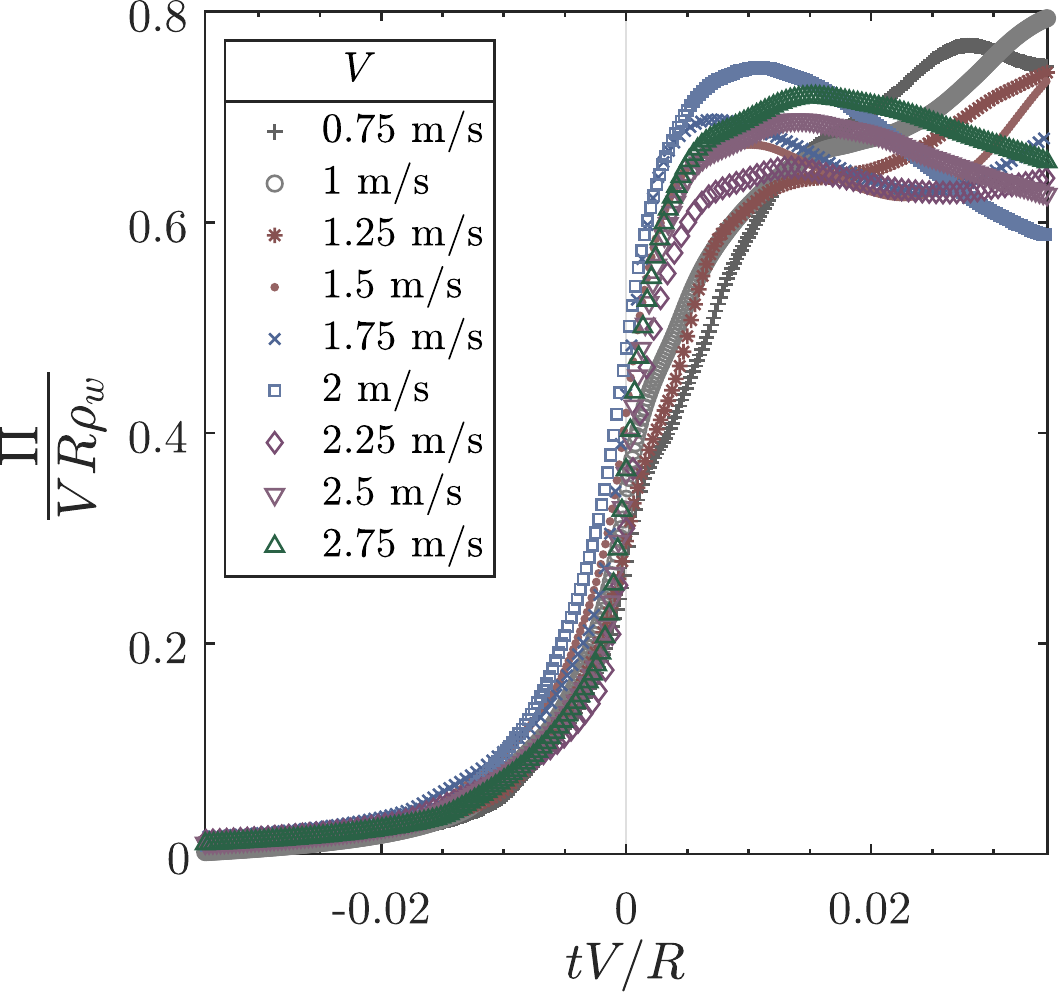}
    \caption{Non-dimensional pressure impulses $\hat{\Pi} = \Pi/V R \rho_w$ measured at disc centre (from figure \ref{fig:pressuresensorimpulsesdim}) are plotted against non-dimensional time $t V /R$. The convincing data collapse {in the region of interest (below $t V/R=0$ in the early stages of impact loading)} reveals that the accumulation of pressure impulse at disc centre, when it impacts at high velocities, can be described as an inertial process.}
    \label{fig:pressuresensorimpulses}
\end{figure}

The air film plays this role of a cushioning layer only in the earliest stages of impact before the peak pressure at disc centre reaches its maximum. Thus, whatever subsidiary effect that is caused by it, such as mitigating pressure singularities, distributing the loading, and prolonging the impact, lasts for a short duration of a few milliseconds prior to $t=0$ in figure \ref{fig:pressuresensorimpulsesdim}. This stage is key to determining the final value of impulse that is accumulated during the first impact peak. {Noting that inertially, pressure should scale as $\rho_w V^2$, and time as $R/V$, we expect the pressure impulse $\Pi(t)$ to scale as the product of the former. Using this inertial scaling,} we non-dimensionalise the results in figure \ref{fig:pressuresensorimpulsesdim}(a) \dev{together with data for other values of $V$} and re-plot them in figure \ref{fig:pressuresensorimpulses}. The growth of \dev{the} pressure impulse at the centre \dev{prior to the peak at $t=0$ s} is very well collapsed by the non-dimensionalisation. {Thus, similar to the growth of stagnation pressure on the free surface under the disc's centre before impact, the \dev{air-mediated} early-stage loading on the disc centre is an inertial process.} This allows us to conclude that at the disc centre, {the impact impulse arises mainly} due to sudden acceleration of a certain bulk of liquid by the impact. {It} leads us to expect that the impulse created by total force on the discs' entire surface should behave similarly to the pressure impulses as in figure \ref{fig:pressuresensorimpulses}. This indicates that early-time loading on the pressure sensor grows as it would have, had the pressure sensor already been in touch with, or immersed in the liquid. {This finding is in keeping with those of \citet{rosshicks2019}, who found using a viscous air-cushioning analysis of a flat-bottomed wedge, that the initially trapped volume of air moves with the same velocity as the impactor. In the added-mass sense for the surrounding (target) liquid, this trapped cushioning layer essentially acts as \dev{an} extension of the solid body.}

\dev{In fact, if for our disc impact experiment one estimates the time it takes for the entrapped air pocket to reach pressures of the order of the stagnation pressure in the liquid, after which the liquid will be able to start moving, one obtains values smaller than $10 \mu$s (see appendix~\ref{sec:pressurisation}). Since this pressurisation time is much shorter than the build-up time of the load, the latter will be dominated by what is happening in the liquid, which connects to the inertial scaling that was observed.}

\section{Spatially integrated loading}\label{sec:loadcellsection}
\subsection{Force measurements and impulse calculation}
\begin{figure}
    \centering
    \includegraphics[width=.99\linewidth]{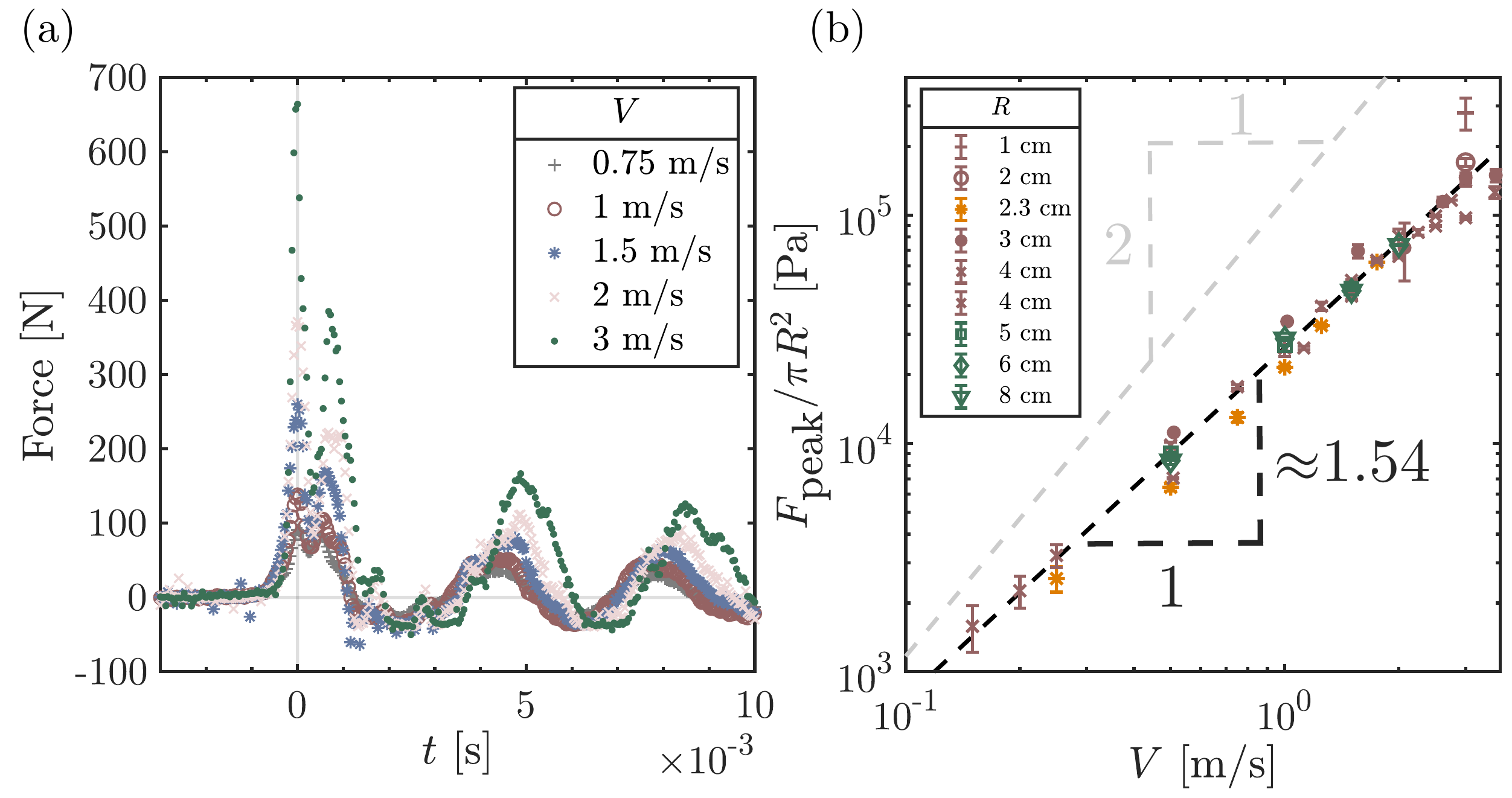}
    \caption{(a) Time evolution of the impact force measured using the disc shown in figure \ref{fig:forceimpulsesetup}(b), impacted at various velocities. The time series are centred about $t=0$ at the time of registering peak impact force. {The natural frequencies of the impacting structure used here were measured with the respective load cells. They ranged from 313.2 $\pm$ 55.5 Hz for the lightest disc, to 236.8 $\pm$ 30.5 Hz for the heaviest attachment.} (b) Peak average pressure, $\tilde{P}_{\text{peak}} = F_{\text{peak}}/\pi R^2$ computed from the force measured in experiments using discs of varying sizes, materials and impact speeds are plotted against their velocities of impact. Data shown using purple coloured markers are from experiments impacting steel discs, yellow from impacting 3D printed plastic discs, and green markers from experiments using aluminium discs. Peak average pressures are found to deviate substantially from the previously found quadratic dependence on $V$ for the centre and (to a lesser extent) the side pressures presented in figure \ref{fig:peakpressuresquadscaling}(c).}
    \label{fig:force1}
\end{figure}

We perform force measurements on the discs to check to what extent they are consistent with the pressure measurements. Some typical force measurements are shown in figure \ref{fig:force1}(a). The main features of interest around the impact peak are the same as were found in pressure measurements. The first large peak results from the initial impact. Here however, the forces measured by the load cell are averaged throughout the disc \& rod assembly that is mounted on the linear motor. As a result, structural vibrations in the impactor dominate the measured signals shortly after $t=0$, and drown several weaker signals that might be of interest.

Force measurements were done using discs of different sizes and materials. The variation of peak forces with $V$ is shown in figure \ref{fig:force1}(b). As with pressure measurements, force measurements at each set of parameters were repeated a number of times to obtain variance in the measurements. They are found to vary substantially from the quadratic scaling that peak pressures were found to follow. Instead, peak forces appear to grow as $\sim V^{1.54}$ within the experimental range of errors. Since experiments done with discs made of steel, aluminium, and plastic follow a consistent trend, we can conclude that the total force produced on the discs indeed have a hydrodynamic origin. They do not depend on the material that a disc is made from. The peak {pressures}' deviation from a $V^2$ scaling appear likely due to limited temporal resolution of the strain gauge load cell.

{However, the challenge in resolving the peak in flat plate impacts is more indicative of the very short time scales involved in this particular impact scenario, rather than only the limited resolution of the equipment. For objects such as a wedge, a cone, or a sphere entering water, the wetted surface expands in time. The impact force increases with the increase in wetted area. Space-averaged \citet{wagner1932stoss} and Logvinovich \citep{korobkin_2004} models for the prediction of force on a water-entering object with a deadrise angle take into account the wetting rate to predict the impact force's time evolution. Force measurements by \citet{vincent_xiao_yohann_jung_kanso_2018} show how increasing the deadrise angle of the wedge increases the duration of the impact peak. \citet{ELMALKIALAOUI2012183} also find well resolved force measurements in impacting cones with deadrise angles varying between $7-30\degree$ while using an accelerometer with a natural frequency of 2500 Hz, and compared successfully with calculations by \citet{shiffman1951force}. Similarly impact drag coefficients measured by \citet{moghisi_squire_1981} in the liquid-entry of a sphere are both well resolved and compared successfully with calculations by \citet{shiffmanspencer1945}. Another indication of how the timescales at first impact peak are affected by the curvature of the impacting face was provided by \citet{ermanyuk2011} by performing slamming experiments with discs with flat, convex and concave surfaces. \citet{ermanyuk2011} associate the greatest accelerations, and correspondingly the shortest timescales in the pressure timeseries, with the water impact of a disc with flat surface, followed by the convex and concave surfaces respectively. Finally we comment that ignoring air-cushioning effects, a finite curvature of the impacting surface necessarily dilates the duration of the initial added-mass induced force peak.}

\begin{figure}
    \centering
    \includegraphics[width=.8\linewidth]{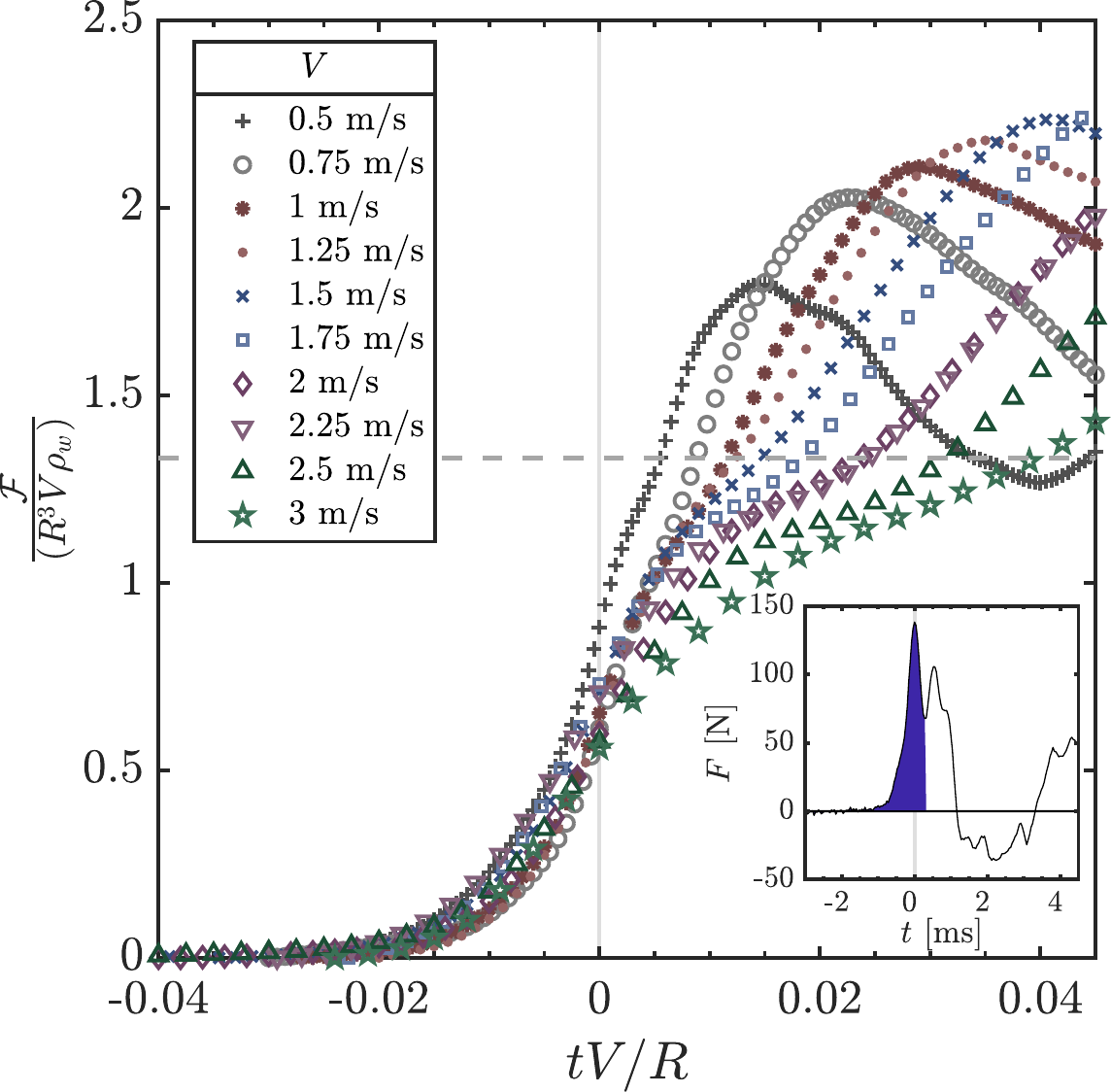}
    \caption{Non-dimensional force impulses $\mathcal{F}/R^3 V \rho_w$ are computed from force measurements with a steel disc of radius $R =$ 4 cm, impacting on water at various $V$. Each result is centred about $t=0$ at the time of the when the corresponding impact force attains its peak value. The horizontal dashed line is drawn to mark the non-dimensional force impulse value of 4/3. Inset: A force-time curve at $V=1$ m/s is shown; the shaded area shows the time-duration over which $F(t)$ is integrated to obtain $\mathcal{F}$.}\label{fig:forceimpulse_steel}
\end{figure}

{However, coming back to the present experiments, and} notwithstanding the limited resolution of load cell, we can numerically integrate the force measurements starting from the time when the first impact peak starts to rise. The force impulses \begin{equation}
    \mathcal{F}(t) = \displaystyle\int_{t_0}^{t} F(t') dt',
\end{equation} thus computed are non-dimensionalised as before using inertial length and time scales. The results are plotted in figure \ref{fig:forceimpulse_steel}. As with pressure impulses, the growth of force impulses until the time of peak force (conveniently at $t=0$) is well collapsed by the re-scaling. Our expectation from previous section (based on figure \ref{fig:pressuresensorimpulses}) is upheld by force impulse measurements.

After $t=0$, the structural oscillations (signals with a period of 4 ms, visible both in figures \ref{fig:peakpressuresquadscaling}(a--b) and \ref{fig:force1}(a)) become very large, and depending on its extent of submergence in the bath, are variedly affected by the bath's sloshing. Notice also in the inset of figure \ref{fig:forceimpulse_steel}, that there is a secondary peak in the force, which cannot be easily ascribed to any hydrodynamic phenomenon. This explains why the later force signals and their resulting impulses do not follow hydrodynamic scales.

\subsection{Impulse due to water impact of a disc}
That at least in our experimental setup, both the pressure impulses $\Pi(t)$, and force impulses $\mathcal{F}(t)$ obey inertial scaling despite the presence of cushioning air layer, is an important observation as regards to comparing the measurements with theory. Here, we start with a brief discussion of the well-known idea of pressure impulse in a fluid. The idea is useful when fluid boundaries, or a portion of the bulk, is subjected to a sudden acceleration of a large magnitude and short duration. Over this short duration, velocities of both the fluid and the boundary are small. Ignoring viscosity, Euler's equation, linearised in $\mathbf{v}$, can be used to describe fluid motion \citep{cooker1995pressure} \begin{equation}
    \frac{\partial \mathbf{v} }{\partial t} = -\frac{1}{\rho_{{w}}} \Vec{\nabla} p,
\end{equation}
where both the terms are manifestly of large magnitude for a short interval of time, in which the acceleration term is dominant over the terms in Euler equation. Integrating both sides with time,\begin{equation}\label{eqn:impulsevelocity}
    \mathbf{v}_{\text{after}} -  \mathbf{v}_{\text{before}} = -\frac{1}{\rho_{{w}}} \Vec{\nabla} \left( \displaystyle\int_{t_\text{before}}^{t_\text{after}} p(t') \mathrm{d}t' \right),
\end{equation}
which can be integrated and re-written in terms of a scalar flow potential ($\Vec{v} = \Vec{\nabla} \phi$) as \begin{align}
    \phi_\text{after} - \phi_\text{before}   = \frac{\Pi(t_{\text{before}}) - \Pi(t_{\text{after}})}{\rho_w}.
\end{align}
This expresses that in a fluid that is initially stationary, any motion produced by sudden acceleration will, at least for a short duration, be irrotational. In such a fluid domain, flow potentials throughout the bulk {are} uniquely determined by the normal component of velocities imposed at its boundaries. The pressure impulse calculation that corresponds exactly to the present experiment with the impact of a disc shown in \citet[section 6.10]{batchelor1967introduction}, {yielding $\mathcal{F}_{\text{impact}} = 4/3R^3V\rho_{{w}}$}. The same result was also arrived at in \citet[section 102]{lamb1945hydrodynamics} by deriving the kinetic energy that is imparted to the surrounding fluid by an immersed disc that is impulsively accelerated from rest. \citet{glasheen1996vertical} performed experiments impacting a disc on water, and measured the loss in its momentum from impact by measuring the sudden drop in its velocity (equation \eqref{eqn:impulsevelocity}). From the loss in the projectile's momentum, they estimated the added mass on the moving disc. The results were compared to the above analytical {result} (also in \citet{birkhoff1957jets}), finding a good agreement.

Note that our method of computing the impact impulse differs significantly from that of \citet{glasheen1996vertical}, who could calculate the force impulse by measuring the sharp drop in projectile momentum. Their method would not be suited in our case due to the amount of control in our experiment that allows us to impose a constant velocity. In contrast, we numerically integrate the directly measured force and local pressure over the duration of the impact peak.

{Although in essence both the methods to measure $F$ relate to the added mass coefficient being measured, the difference between the two approaches can be understood by using the following expression for the inertial force on the disc in terms of the depth of penetration $h(t)$ into the liquid phase as derived by \citet{iafrati2011asymptotic}
\begin{eqnarray}
F &=& m_a \rho_w R^3 \ddot{h} + \rho_w {R^2} \dot{h}^2\left[a_1\left(\frac{h}{R}\right)^{-1/3}  + a_2 \log \left(\frac{h}{R}\right) + a_0   \right] + O(1)\nonumber\\
 &\approx& m_a \rho_w R^3 \ddot{h} + a_1\rho_w {R^2} \dot{h}^2 \left(\frac{h}{R}\right)^{-1/3}\,,
\label{Iafrati1}
\end{eqnarray}
where $\dot{h}$ and $\ddot{h}$ denote the first and second time derivative of $h(t)$. Here, the first term represents the added mass (with $m_a = 4/3$) and the second term the building up of quadratic drag, within which the first contribution (with $a_1 \approx 2.7250$) is the dominant one for small $h$. Note that the quadratic drag term is diverging and therefore dominant for small $t$. We may time-integrate the dominant contribution to Eq.~\eqref{Iafrati1} from a moment $t=0^{-}$ just before impact, where $\dot{h}(0^-) = 0$, to some time $t=t_1$ after impact, which yields the impulse
\begin{eqnarray}
\mathcal{F} &=& \int_{t = 0^-}^{t_1} Fdt \approx \rho_w R^3 \left[m_a\dot{h}(t_1)  + \tfrac{3}{2} a_1\int_{t = 0^-}^{t_1} \dot{h}\,d\!\left((h/R)^{2/3}\right)\right]\,.
\label{Iafrati2}
\end{eqnarray}
One may estimate the second term by noting it is always positive and, since $\dot{h} \leq V$ we have
\begin{equation}
\tfrac{3}{2} a_1\int_{t = 0^-}^{t_1} \dot{h}\,d\!\left((h/R)^{2/3}\right) \leq\tfrac{3}{2}a_1 V \int_{t = 0^-}^{t_1} d\!\left((h/R)^{2/3}\right) = \tfrac{3}{2}a_1\left(\frac{h(t_1)}{R}\right)^{2/3}\,,
\label{Iafrati3}
\end{equation}
that is, in contrast to its dominance in the expression for the force for small $t$ (Eq.~\eqref{Iafrati1}), the contribution of the quadratic drag term to the impulse is vanishingly small for small $h$, and the impulse is dominated by the added-mass contribution, $\mathcal{F} \approx m_a \rho_wR^3 \dot{h}(t_1)$.}

{Clearly, when the velocity is kept constant, we find that $\mathcal{F} \approx m_a \rho_wR^3 V$ in the inertial approximation, and one would expect the impulse to quickly converge to $m_a \rho_wR^3 V$. Any deviation would be due to secondary effects, such as liquid compressibility which accounts for the growth of the added mass and takes place on a time scale $R/c_w$ with $c_w \approx 1.5\cdot10^3$ m/s the speed of sound in water, which in our experiments is of the order of several tens of microseconds.}

{However, when the velocity is not controlled, e.g., when the impactor has a mass $M$, then the impulse is not constant since $h(t)$  is the solution of the equation of $M\ddot{h}  = - F + Mg$, where $F$ is given by Eq.~\eqref{Iafrati1}. Even if we neglect the acceleration due to gravity $g$, this implies that $\dot{h}(t)$ is changing from its value upon impact ($V$) to its terminal value ($V/(1+m_a/M)$) on a time scale that is determined by the quadratic drag term, which is necessarily proportional to the inertial time $R/V$, which would be of the order of 10-100 ms for experiments similar to ours. During this time interval the force that the disc experiences is determined by the quadratic drag term in an indirect manner, namely by how it affects $h(t)$, and consequently $\dot{h}$ and $\ddot{h}$.}

{Thus, there is a crucial difference in free fall experiments reported before, and our experiments imposing a constant velocity. By keeping a constant velocity on the disc, we effectively suppress effects that cause variation of the added mass terms in equations \eqref{Iafrati1} and \eqref{Iafrati2}, thereby bringing the experiments closer to the classical theory. Additionally,} our measurements extend the range of experimental parameters ($R^3 V \rho_{{w}}$) covered by \citet{glasheen1996vertical} to much larger {range of} disc momenta at impact.

\begin{figure}
    \centering
    \includegraphics[width=.65\linewidth]{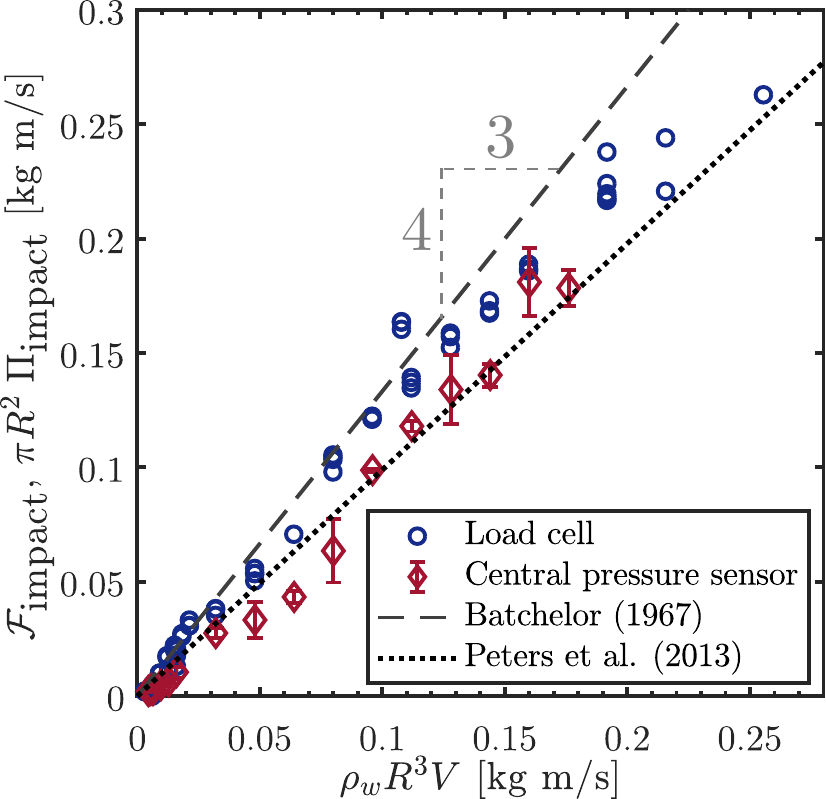}
    \caption{Impulses accumulated during the first impact peak (see inset of figure \ref{fig:forceimpulse_steel}) are plotted against growth rate of the added mass of liquid at impact. Impulses calculated from pressure measurements ($\Pi_{\text{impact}}$) at the disc centre (shown in figure \ref{fig:pressuresensorimpulses}) were multiplied by the disc area so that they can be directly on the same scale as $\mathcal{F}_{\text{impact}}$. Dashed line shows the theoretical calculation for the impulse transferred from an impulsively accelerated disc to an infinite half-space filled with inviscid fluid. Dotted line is the result from \citet{peters2013splash} (equation \eqref{eqn:petersimpulse}).}
    \label{fig:forceimpulse43}
\end{figure}

We compute $\mathcal{F}_{\text{impact}}$ and $\Pi_{\text{impact}}$ by considering the signal upto the end of the first peak (as shown in inset of figure \ref{fig:forceimpulse_steel}). {$\Pi_{\text{impact}}$} are multiplied by the disc area to make the two measures of impact impulses comparable. The comparison is shown in figure \ref{fig:forceimpulse43}. Note that the range of data for $\mathcal{F}_{\text{impact}}$ is greater than that for $ \pi R^2  \Pi_{\text{impact}} $. This is due to the forces being measured with disc sizes up to $R = 8$ cm, while the pressures were measured with fixed disc size of $R=4$ cm. It is seen in figure \ref{fig:forceimpulse43} that force impulses from our measurements lie very close to the theoretical prediction. We find good agreement between our data and the $4/3$ slope for small impact momentum ($\rho_w R^3 V$). However at larger momenta, discrepancies set in, which cannot be solely attributed to limited temporal resolution of the sensors. {We also do not anticipate any effects from finite depth of the bath to have played in our measurements. Experiments by \citet{ermanyuk2011} showed the effects of finite bath approaching the asymptotic value of added mass coefficient (equalling 4/3, also see \citet{CHEBAKOV1974628}) for a bath of depth $0.8 R$. Bath depths in the present work ranged between $3.75R-30 R$.}

It is possible to at least approximately compute the influence of entrapped air layer on $\mathcal{F}_{\text{impact}}$, as was done by \citet{peters2013splash}, who calculated the total force on a disc with the cushioning air layer. They estimated the change in momentum of a disc slamming at constant velocity as \begin{equation}\label{eqn:petersimpulse}
   \mathcal{F}_{\text{impact}} = 0.315 \rho_{w} \pi R^3 V,
\end{equation}
i.e., a reduction of the classical force impulse by about 25\%. This result is also compared to our measurements in figure \ref{fig:forceimpulse43}. It lies closer to the area-integrated pressure impulses measured at the disc centre, but consistently below the directly obtained force-impulse data from the load cell.

\section{Conclusions}
We report here experiments where a flat disc is impacting at a controlled, constant velocity on a deep water bath initially at rest. We measure local pressures at the center and near to the edge of the disc, and the total impact force. At the moment of impact the disc entraps a thin air layer on the impacting side due to air-cushioning that occurs prior to first touchdown. The air layer causes a difference in impact pressures at the disc's centre and edge. {The first pressure peak is registered at the disc edge, where \dev{the disc} makes first contact with the liquid. While} {the air layer} prolongs the {build-up of the pressure peak}, {we also find that the time duration between the occurrence of pressure peaks at the two sensors is always of \dev{the order of 100 $\mu$s}
regardless of how the air film behaves. The trapped air film retracts or collapses over longer time scales ($\sim \mathcal{O}\text{ } 10^{-1} \text{ s}$ and $10^{-3} \text{ s}$ respectively), \dev{which implies that} 
the film's time evolution is much slower \dev{than the buildup of the load}. The pressure peak at the centre has already occurred before the shape of entrapped air film has changed significantly.}

Pressure impulses $\Pi(t)$ are computed by numerically integrating \dev{pressure measurements} starting from \dev{a time before} they start to rise. The pressure impulses $\Pi(t)$ accumulated at the disc centre are found to be higher than those at the disc edge. Further, the early\dev{, air-mediated} growth stage of the central pressure impulse is {shown} to be governed by inertial length and timescales\dev{, which may be connected to the very fast pressurisation of the air layer, after which the dynamics is expected to be dominated by what happens in the liquid.}

\dev{By multiplying with the disc area,} the peak pressures are compared with peak forces on the disc. The peak forces are found to not follow $V^2$ scaling, {which is only indicative of insufficient resolution of the force sensor. The impulses, being a more practical measure of the intensity of loading, are used to isolate what role the air layer in cushioning of the impact loading.} The role of added mass in producing the large initial impact force peak is verified by determining the force impulse $\mathcal{F}(t)$ at impact. As seen with $\Pi(t)$, the growth of $\mathcal{F}(t)$ {in the early stages of loading} is also well collapsed by a re-scaling with inertial length and time scales. Our analysis{, in line with findings from \citet{bagnold1939interim}, \citet{hattori1994wave} and \citet{partenscky1989dynamic}}, shows that impulses are a much more reliable, and reproducible indicator of the intensity of an impact event. This is especially the case when sensors are suspected to underestimate the peak pressures (or forces) due to their limited time resolution.

The impulses accumulated during the first peak, $\Pi_{\text{impact}}$ and $\mathcal{F}_{\text{impact}}$, are compared to the theoretical hydrodynamic mass of the fluid that is accelerated by the impact. A {better} agreement is found between force impulses and the theory than by \citet{glasheen1996vertical}, {and over a much larger range of parameters}. The $\Pi_{\text{impact}}$ at disc centre are found to be consistently lower than both $\mathcal{F}_{\text{impact}}$ and the \dev{theoretical prediction}. At {larger} disc momenta, the data increasingly deviate from the $4/3$ slope, and follow a trend closer to the {$\mathcal{F}$} calculation by \citet{peters2013splash}{, which was done while accounting for the air cushioning effect}.

\section*{Acknowledgements}
We thank JM Gordillo for useful discussions. We acknowledge financial support from SLING (project number P14-10.1), which is partly financed by the Netherlands Organisation for Scientific Research (NWO). P.V-M acknowledges the support of the Spanish Ministry of Economy and Competitiveness through grants DPI2017-88201-C3-3-R and DPI2018-102829-REDT, partly funded with European funds.

\section*{Declaration of Interests}
The authors report no conflict of interest.

\appendix

\section{\dev{Time evolution of the entrapped air layer}}\label{sec:timeev}

{The \dev{fate} of the trapped \dev{air layer} is determined by a balance of the stagnation pressure \dev{that occurs as the water flows along the air-water interface}, surface tension \dev{which tries to minimize its surface area,} and its natural oscillations. \dev{At low impact velocities, surface tension overcomes inertia and the air bubble retracts.} At high impact velocities, when the inertial time scales are much shorter than those relevant for the retraction (see figure \ref{fig:airfilmcollapse}), the air bubble is rapidly \dev{punctured, fragmented and expelled}. Imperfections present in the experiment, such as slight tilt of the disc, or its surface roughness, also play an increasingly interfering role in triggering a rupture of the air film. As such it becomes difficult to define a clear criteria to separate the two `retracting' and
`fragmenting' behaviours. Nevertheless, it is found for several parameters that the two co-exist, showing that the two are not in fact mutually exclusive `regimes' (see for example, movies 5 and 6 showing water entry of 5 and 9 cm wide discs at 0.5 m/s and 0.3 m/s respectively).}

{Some quantitative discussion on the behaviour of the \dev{entrapped air layer} is presented \dev{in this appendix.}}

\subsection{\dev{Retraction of the air film at low impact velocities}}\label{sec:appairfilmretraction}
\begin{figure}
    \centering
    \includegraphics[width=.6\linewidth]{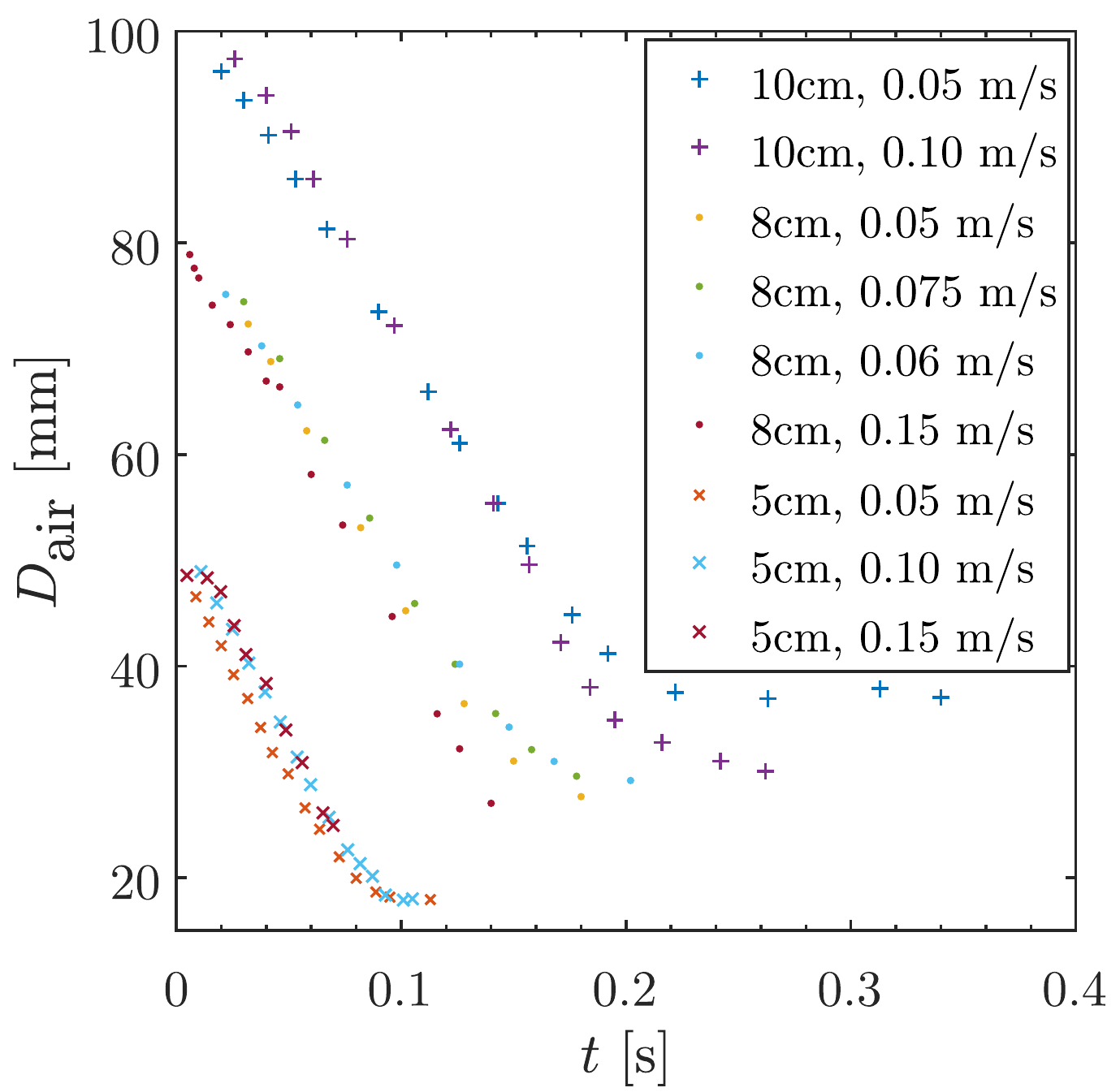}
    \caption{\dev{Time evolution of the diameter $D_\textrm{air}$ of the retracting air film at low impact velocities for different disc diameters $D$ and impact velocities $V$. Note that the retraction velocity $V_\textrm{retr} \approx 0.38-0.42$ m/s appears to be independent of $D$ and $V$.}}
    \label{fig:airfilmretraction}
\end{figure}
{The inwards retraction of the air film at low \dev{impact} velocities \dev{is} due to the surface tension of the \dev{air-water interface. We measured the time evolution of the bubble diamter for different disc diameters and impact velocities and present them in} figure \ref{fig:airfilmretraction}. The retraction velocity $V_\textrm{retr} \approx 0.38-0.42$ m/s appears to be near-constant, suggesting a Taylor-Culick type retraction. Similar observations were made by \citet{mayer2018flat} while using a rectangular plate.}

\dev{It is good to note that the }
{time scale of retraction $\sim \mathcal{O}(10^{-1} \text{ s})$ is significantly larger than that over which \dev{the load on the disc builds up and} the peak pressures at the disc edge and center are separated in time $\sim \mathcal{O}(10^{-4} \text{ s})$ (inset, figure \ref{fig:peakpressuresquadscaling}(b)).}

{\dev{In addition, we observe that the} retracting bubble's perimeter emits surface waves that converge to its center and produce the largest oscillations. An example is shown in Movie 7 (12 cm disc entering water at 0.4 m/s). Waves with a large range of wavelengths are initially excited. Those with the smallest wavelengths travel the fastest. Such a bubble's natural frequency, despite being attached and flattened against a plate, is essentially the same as Minnaert frequency \citep{blue1967}. For $R=6$ the resonance frequency is approximately 54 Hz, with the corresponding oscillation period of approximately 18.5 ms. It can be seen in Movie 7 that by 18.5 ms after the initial touch-down, several oscillation cycles at the center have already occurred at the bubble's center, thereby showing that the bubble's natural oscillations do not affect either its retraction, or its puncturing.}

\subsection{{Puncturing of the air film at \dev{high impact velocities}.}}\label{sec:appairfilmpuncture}
\begin{figure}
    \centering
    \includegraphics[width=.55\linewidth]{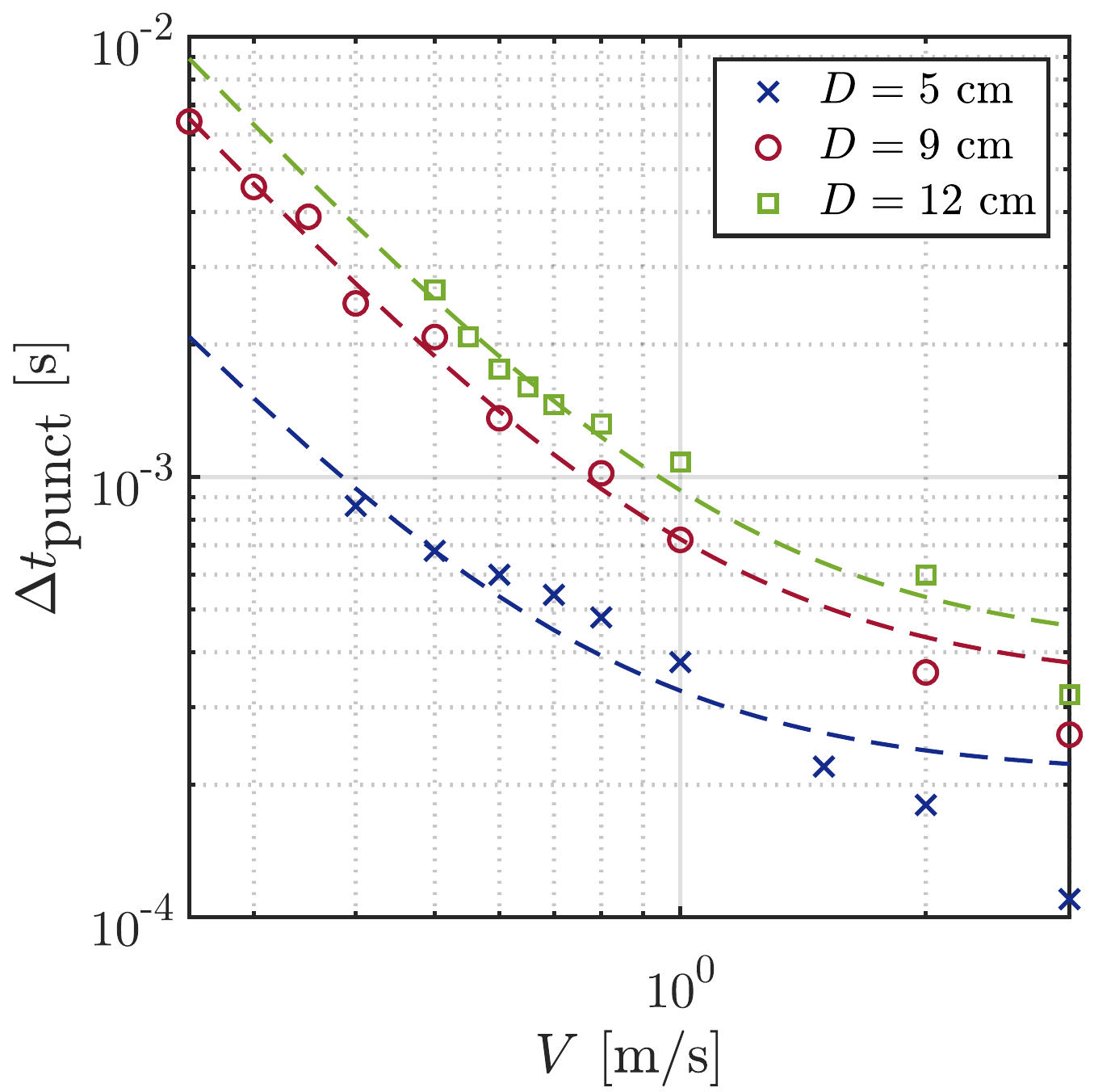}
    \caption{\dev{Time interval $\Delta t_\textrm{punct}$ between the first contact time of the disc and the water surface and the appearance of the puncture in the center of the air film, as a function of the impact velocity $V$ for three different disc diameters. The dashed lines are best fits to the functional form $\Delta t_\textrm{punct} = A/V^2$, with $A$ constant.}}
    \label{fig:onsetofpuncture}
\end{figure}

\dev{The time interval $\Delta t_\textrm{punct}$ between the first contact time of the disc and the water surface and the appearance of the puncture in the center of the air film are measured as a function of the impact velocity $V$ for three different disc diameters and presented in figure \ref{fig:onsetofpuncture}.} {A smaller disc entraps a thinner air layer \citep{ujthesis, jainkhPRF}, with the consequence that the air layer with the smaller discs is more susceptible to being affected by experimental imperfections such as disc surface roughness or its edges. \dev{The time interval $\Delta t_\textrm{punct}$ is reasonably well fitted with the functional form $\sim 1/V^2$,} with the exception of the smallest disc where we do expect the the air layer to be most sensitive to perturbations from experimental imperfections - leading it to rupture earlier than otherwise, or at locations off the centre.} \dev{It is good to note that the time interval $\Delta t_\textrm{punct}$ is of the order of a millisecond, and decreases with the disc diameter $D$.}

{Some examples of the process are also shown in Movies 4--6. Movies 5 and 6 were particularly chosen to show that \dev{puncturing may also occur for a} retracting air film \dev{as evidence that both may occur simultaneously and therefore do not constitute} separated regimes. The two videos using $D=$ 5 and 9 cm, with $V=$ 0.5 and 0.3 m/s, show \dev{how} an initially retracting air film such as described in appendix \ref{sec:appairfilmretraction}, can still eventually rupture \dev{close to the center} at a later stage \dev{in time}.}

\dev{\subsection{Pressurisation of the entrapped air pocket}}\label{sec:pressurisation}

\dev{The last topic we want to discuss is the pressurisation of the air pocket. As soon as the air pocket is entrapped, the disc further moves down into the liquid and air pocket will be compressed. The entrapped air can only begin to displace the surrounding liquid once it has reached a value comparable to the stagnation pressure. To estimate the amount of time needed for this to happen, we assume adiabatic compression such that pressure $p$ and volume $v$ of the entrapped air pocket are related by
\begin{equation}
    p v^\gamma = p_0 v_0^\gamma\,,
\end{equation}
where $p_0$ and $v_0$ are the ambient air pressure and the initial volume of the entrapped air pocket, respectively. For a disc of surface area $S = \pi R^2$, we know from \citet{jainkhPRF} that an air film of thickness $h_0 \approx\kappa R$ is entrapped at impact, where  $\kappa \approx 7 \cdot 10^{-3}$. The initial volume of this air pocket can therefore be estimated as $v_0 \approx S h_0$. On compression in the vertical direction, the decrease in volume $S \Delta h$ relates to an increase in pressure $\Delta p$. For a disc plunging at constant velocity $V$, $\Delta h = V \Delta t$. The adiabatic compression law thus reads
\begin{equation}
    (p_0 + \Delta p) \; S^\gamma (h_0 - V \Delta t)^{\gamma} = p_0 \, S^\gamma h_0^ \gamma,
\end{equation}
from which we can compute the time $\Delta t_c$ it takes to pressurise the air pocket to a pressure $\Delta p$ above $p_0$ \begin{equation}
    \Delta \tau_c \equiv \frac{V \Delta t_c}{R} =  \kappa \left[ 1 - \left( 1 + \frac{\Delta p}{p_0} \right)^{-1/\gamma} \right]\,,
\end{equation}
where $\Delta \tau_c$ is the non-dimensionalised pressurisation time. The time taken for the pressure in the air pocket to rise to the stagnation pressure $\Delta p = \rho V^2$ can now be estimated using $\gamma = 1.4$ for air, $\rho=10^3$ kg/m$^3$ and $p_0 = 10^5$ Pa. }

\dev{In table \ref{table:pressurisationtime}, we calculate the pressurisation time for $R = 1-8$ cm and $V = 0.5 - 3$ m/s which correspond to our slamming impact experiments. From this table we conclude that it takes between $0.3-10$ $\mu$s to pressurise the air pocket to stagnation pressure, after which the pressurised air pocket can begin to displace the surrounding liquid. Clearly, this corresponds to time scales which are very short compared to the build-up time of the load.}

\begin{table}
  \begin{center}
\def~{\hphantom{0}}
{
  \begin{tabular}{l c | c | c | c |}
      $V$ [m/s]  & $\Delta \tau _c$   &   \shortstack{$\Delta t_\text{stag}$ [$\mu$s] \\ $_{R=1 \text{ cm}}$} &  \shortstack{ $\Delta t_\text{stag}$ [$\mu$s]\\  $_{R=8 \text{ cm}}$ }   \\\hline \\[3pt]
       0.5   & $12.6 \cdot 10^{-6}$ &  0.25  & 2.0  \\
       1.0   & $49.6 \cdot 10^{-6}$ & 0.50  & 4.0 \\
       1.5  & $110 \cdot 10^{-6}$ & 0.73 & 5.9 \\
       2.0 & $193 \cdot 10^{-6}$  & 0.97 & 7.7 \\
       3.0 & $418 \cdot 10^{-6}$ &  1.39 & 11.1 \\
  \end{tabular}
  \caption{Estimates of the compression time $\Delta t_c$ for $R= 1$ and $8$ cm, with $V = 0.5 - 3$ m/s.}}
  \label{table:pressurisationtime}
  \end{center}
\end{table}

\bibliographystyle{jfm}
\bibliography{bibliography}

\end{document}